\documentclass[aps,prl,twocolumn,superscriptaddress,amsfont,graphicx,nofootinbib,preprintnumbers]{revtex4}%
\usepackage{color,graphicx,epsfig}
\usepackage{ifpdf}
\usepackage{amsmath}
\usepackage{bm}
\usepackage{color}
\usepackage[english]{babel}
\usepackage{graphicx}%
\usepackage{amsfonts}%
\usepackage{amssymb}
\usepackage{braket}
\usepackage{hyperref}
\usepackage{enumerate}
\definecolor{nicered}{rgb}{0.7,0.1,0.1}
\definecolor{nicegreen}{rgb}{0.1,0.5,0.1}
\hypersetup{colorlinks,citecolor= nicegreen,linkcolor= nicered}

\newcommand{\beq}{\begin{equation}}
\newcommand{\eeq}{\end{equation}}
\newcommand{\bea}{\begin{eqnarray}}
\newcommand{\eea}{\end{eqnarray}}

\definecolor{Red}{rgb}{1.,0.,0.}

\def\ah{$\Delta \alpha_{\rm{had}}$}

\def\muone{MUonE}
\def\OMIT#1{}

\begin{document}

\def\Buffalo{Department of Physics, University at Buffalo \\ The State University of New York, Buffalo 14260 USA}

\preprint{
}
\flushbottom

\title{The interplay between SM precision, BSM physics and measurements of $\alpha_{\rm{had}}$ in $\mu$-$e$ scattering }

\author{Ulrich Schubert}     
\email[Electronic address:]{ulrichsc@buffalo.edu}
\affiliation{\Buffalo}

\author{Ciaran Williams}     
\email[Electronic address:]{ciaranwi@buffalo.edu}
\affiliation{\Buffalo}

\date{\today}
\begin{abstract}
{Muon electron scattering experiments such as the proposed \muone~experiment, offer an opportunity 
for an improved measurement of the Leading Order hadronic running of $\alpha$, denoted \ah. Such a measurement could be utilized to 
reduce the theoretical uncertainty on the prediction of the anomalous magnetic moment of the muon, $g-2$. Currently there  
is a discrepancy between theory and data for this observable which could potentially be explained by Beyond the Standard Model (BSM) physics. 
Here we investigate the possible impact of missing Standard Model (SM) higher order corrections and BSM physics on the proposed measurement of \ah. 
In principle either could be indirectly fitted into \ah, causing inconsistencies if used in a $g-2$ application. 
The literature suggests a target of 10 ppm on the cross section for the theoretical accuracy. We assess the validity of this target in detail using a variety of methods, 
finding that a 1 ppm target is a more conservative estimate to ensure missing higher orders do not dominate the theoretical uncertainty.  
For the potential BSM contributions we study various models which contribute first
at tree- and loop-level. 
Of particular interest is the impact from dark photon models, which can potentially affect the measurement of \ah~at the desired accuracy. 
At loop-level there exists  in general a kinematic suppression adequate to reduce the BSM contributions
to a level which can be neglected for the extraction of \ah. 
}
\end{abstract}

\maketitle

\section{Introduction}

The quest to conclusively establish the nature of physics Beyond the Standard Model (BSM) has driven high energy physics for several decades. 
Extensions to the Standard Model (SM) are well motivated, since the SM lacks a suitable dark matter candidate, as well as a description of gravity, and has some unappealing features, for instance in relation to the hierarchy problem.  
However, recent results from collider experiments (predominantly the Large Hadron Collider (LHC))  paint a picture which is remarkably consistent with 
the predictions of the SM. Barring any major surprises in the current LHC Run II data set, the quest to derail the SM will enter into a precision regime. That is, 
BSM physics will be hunted not through searches for direct production of new particles, but through subtle deviations made manifest in the coupling of SM particles to each other and themselves. 

Excitingly precision tests already put the SM under significant tension. There has been a long standing deviation between the prediction from the SM for the anomalous magnetic momentum of the muon $g-2$, and various experimental measurements of the same quantity. Excitement is building for the upcoming update from the Muon $g-2$ experiment at Fermilab~\cite{Chapelain:2017syu}, which will present first results this summer. The Fermilab experiment should be able to improve upon the current measurement from Brookhaven National Laboratory (BNL)~\cite{Bennett2006}, ultimately aiming to make the experimental uncertainties small enough to claim a five standard deviation discrepancy with the SM.  A challenge in making such a monumental statement is that one must attempt to quantify the theoretical uncertainty in a robust way so as to ensure the validity of the comparison. There is no 100\% infallible method of estimating theoretical uncertainties in the SM. This is particularly the case for calculations of $g-2$, which rely on a delicate mixture of perturbative and non-perturbative ingredients. 
Nevertheless, several independent calculations and methodologies have been performed resulting in predictions which all agree within one sigma~\cite{Davier:2017zfy,Jegerlehner:2017zsb,Keshavarzi2018}, with the most recent result corresponding to 
$a_{\mu}^{SM}= 11659182.04 \pm 3.56 \times 10^{-10}$. Compared to this prediction the current (BNL) observation is 3.7 standard deviations different $a_{\mu}^{SM}= 11659209.1 \pm (5.4) (3.3) \times 10^{-10}$. For a review of the theoretical predictions for $g-2$ we refer the reader to Ref.~\cite{Jegerlehner2009}, and a recent review of potential BSM explanations can be found in Ref.~\cite{Lindner2018}.

While the perturbative piece of $g-2$ is under very good control~\cite{Aoyama:2012wj,Laporta:2017okg,Czarnecki:2002nt,Aoyama:2012wk,Gnendiger:2013pva}, the non-perturbative components are the largest contributors to the 
theoretical error budget. Currently the Leading Order (LO) hadronic contributions and light-by-light scattering dominate the uncertainty, broadly speaking, both contribute approximately 3$\times 10^{-10}$ to the total error estimate~\cite{Jegerlehner:2017zsb,Davier:2017zfy,Keshavarzi2018,Jegerlehner:2017lbd,Prades:2009tw,Melnikov:2003xd,Knecht:2002hr}. 

The most precise predictions for the LO hadronic contributions are currently extracted from the ratio $R = \sigma(e^+e^-\rightarrow {\rm{hadrons}})/\sigma(e^+e^-\rightarrow \mu^+\mu^-)$ coupled with the optical theorem. This extraction is made difficult by the copious amount of low energy QCD bound states~\cite{Colangelo:2014pva,Keshavarzi2018}, which have to be integrated over. 
Efforts are underway to improve the situation. On the one hand Lattice QCD provides means to calculate the hadronic contributions independently from experimental data~\cite{Blum:2014oka,Green:2015sra,Blum:2016lnc,DellaMorte:2017dyu,Borsanyi2018,Blum2018,Giusti2019,Shintani2019,Davies2019,Gerardin2019}. On the other hand to avoid the complications from bound states a new measurement of \ah, in an alternate kinematic regime, has been proposed~\cite{Abbiendi2017,Abbiendi:2677471}, known as the \muone~experiment. 

\begin{center}
\begin{figure}
\includegraphics[width=5cm]{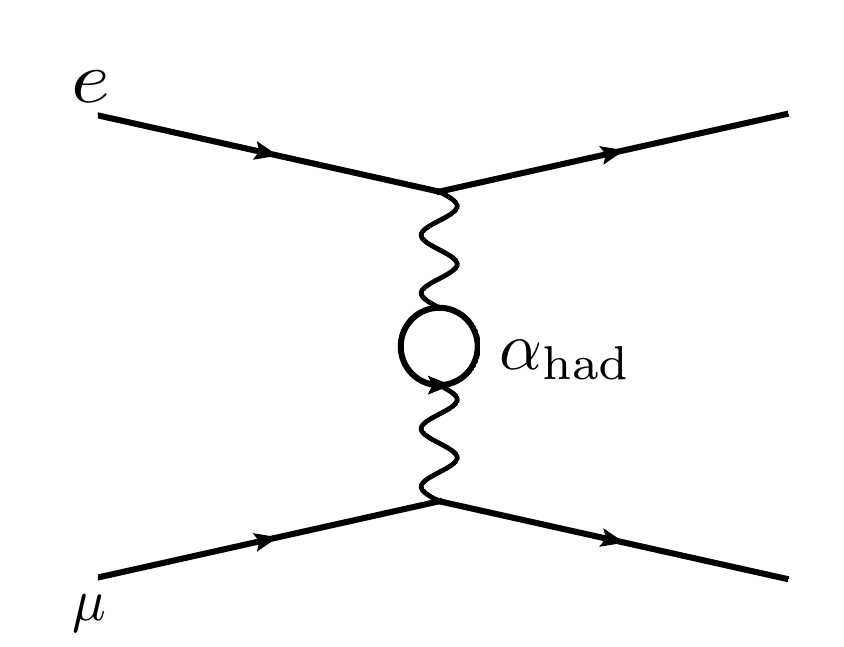}
\caption{Feynman diagram corresponding to the ``signal" for the \muone~experiment}
\label{fig:alph}
\end{figure}
\end{center}

This experiment plans to use a precision measurement of low energy $\mu e$ scattering to probe the running of $\alpha$ as shown in Fig.~\ref{fig:alph}. 
Since the running is now probed in the spacelike $t$-channel regime the integrand is a smooth function and no longer suffers from the complications due to the production (and decay) of QCD hadrons.  
However, the extraction of such an accurate measurement of \ah~ (corresponding to an uncertainty of around 0.3\%~\cite{Dainese:2019xrz}) represents a significant experimental and theoretical challenge. In order to obtain the theoretical accuracy needed, a dedicated effort to provide differential calculations and Monte Carlo codes has begun. In particular the Next-to-Leading Order (NLO) QED and EW effects have been calculated~\cite{Alacevich2019} as well as the NLO and Next-to-Next-to-Leading Order (NNLO) hadronic contributions~\cite{Fael:2019nsf}. Furthermore significant progress has been made towards a full NNLO QED calculation~\cite{Mastrolia2017,DiVita2018}.

While significant attention has been given to the predictions for $\mu e$ scattering in the SM, thus far, to the best of our knowledge, no study has been performed which investigates the sensitivity of \muone~ to BSM physics. That is to say, if BSM physics exists and contributes around $\Delta a_{\mu} = 20 \times 10^{-10}$ to $g-2$, what is the subsequent impact on a scattering experiment (also involving muons) which seeks to measure the hadronic contributions at the level of $2 \times 10^{-10}$? One may naturally worry that any BSM contribution could be present in both to such an extent as to invalidate the methodology. In the worst case scenario, BSM physics would be fitted into \ah~ and the agreement between the ``SM" and data would be artificially enhanced. This paper aims to answer this question.
In order to do so, we will study situations in which BSM enters at both tree- and loop-level and classify the overall impact in a reasonably broad and model independent manner. 
Before doing so we will first reassess the impact of theoretical uncertainties from the SM itself and compare them to the targeted accuracy of $2 \times 10^{-10}$. 

\section{Overview}
\label{sec:overview}

\begin{center}
\begin{figure}
\includegraphics[width=8cm]{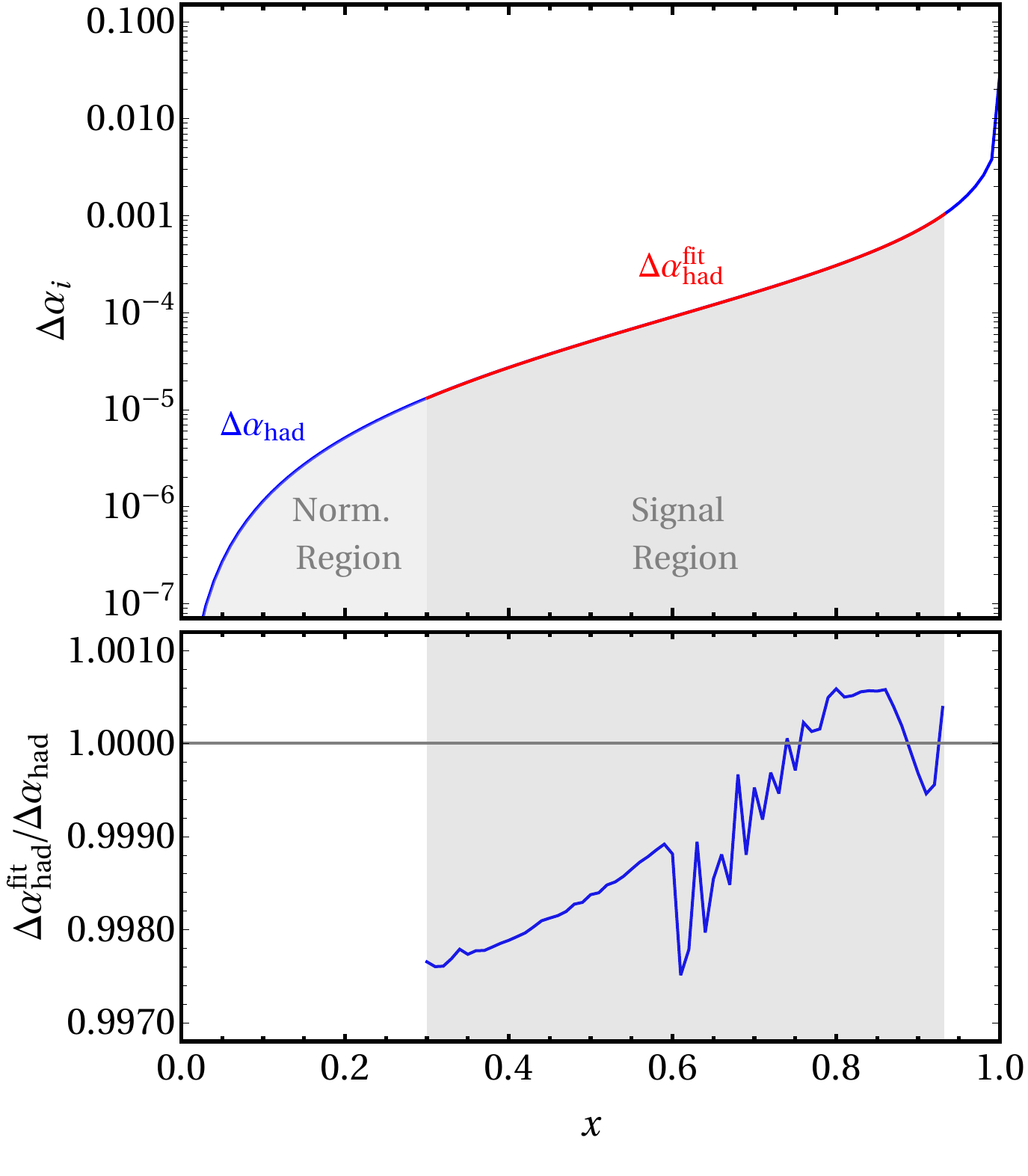}
\caption{The upper panel shows the Leading Order hadronic contributions \ah~to the running of $\alpha$ in the \muone~signal and normalization region (computed using the {\tt{hadr5n12}} program~\cite{Eidelman:1995ny,Jegerlehner:2008zz,Jegerlehner:2017zsb}) and in red we show a cubic fit to \ah~in the signal region. The lower panel shows the ratio of the cubic fit over \ah.}
\label{fig:ahahlo}
\end{figure}
\end{center}

The \muone~experiment proposes to measure \ah~ via $t$-channel scattering of muons and electrons. Once \ah~ is defined, its subsequent contribution to $g-2$, denoted $a^{\text{HLO}}$, is obtained via the following integration 
\begin{eqnarray}
a^{\text{HLO}}=\frac{\alpha}{\pi} \int_0^1 dx (1-x) \Delta \alpha_{\text{had}}[t(x)] \, ,
\label{eq:HL0}
\end{eqnarray}
where $t$ is defined in the spacelike region as a function of $x$ in the following way
\begin{eqnarray}
t(x)=\frac{x^2m_\mu^2}{x-1}<0. 
\end{eqnarray}
In this work we will use the {\tt{hadr5n12}} program~\cite{Eidelman:1995ny,Jegerlehner:2008zz,Jegerlehner:2017zsb}\footnote{specifically the 09/09/2009 version} to generate \ah~ (and subsequently $a^{\text{HLO}}$ via Eq.~\eqref{eq:HL0}), the results from the code for \ah~ as a function of $x$ are shown in Fig.~\ref{fig:ahahlo}, where we have highlighted the \muone~signal region, which corresponds to $x \in [0.3,0.932]$. The region $x < 0.3$ is an area of phase space in which the contribution from \ah~ is rather small, and thus is proposed as a normalization window to aid in the reduction of experimental systematic uncertainties. The region $x > 0.932$ is not kinematically accessible for the proposed experiment. In the signal region \ah~ is well modeled by a cubic polynomial $\Delta \alpha^{\rm{fit}}_{\rm{had}}=c_1\, t + c_2 \, t^2 + c_3 \, t^3$, which is also displayed in the figure. We note that \ah$(0)=c_0 =0$, as such there is no constant term in the fit.   
Over the \muone~ data range, with around 30 data points and 3 free parameters it is possible to obtain fitting errors on $a^{\text{HLO}}$ at the level of 0.3\%~\cite{CernTalk}. 

We can relate the extraction of \ah~ to the perturbative expansion in theory. The differential cross section (in $t$) expanded to NLO in $\alpha$ can be written as 
\begin{eqnarray}
\frac{ d \sigma^{\rm{SM}}_{\rm{full}}}{d t}  &=& \alpha^2 \frac{ d \sigma^{\rm{LO}}}{d t}  + \alpha^3  \frac{d \sigma^{\rm{NLO}}}{d t}  + 2 \alpha^2 \frac{d}{{d t} }\left(\sigma^{\rm{LO}} \Delta \alpha_{\rm{had}}\right)\, , \nonumber\\
&=&\frac{ d \sigma^{\rm{SM}}_{\rm{pert}}}{d t}+ 2 \alpha^2 \frac{d}{{d t} }\left(\sigma^{\rm{LO}} \Delta \alpha_{\rm{had}}\right) \, .
\end{eqnarray}
The first two terms can be readily computed in perturbation theory and are therefore considered a ``background'' in the \muone~ setup, which will be subtracted from the data. 
Access to \ah~ can be found by equating the following ratio  
\begin{eqnarray}
\frac{d}{dt}\left(\frac{d \sigma_{{\rm{Exp}}}} {d \sigma^{\rm{SM}}_{\rm{pert}}}\right)   
= \frac{d}{dt}\left(\frac{d \sigma^{\rm{SM}}_{\rm{full}}}{{d \sigma^{\rm{SM}}_{\rm{pert}}}}\right) \, , 
\label{eq:ExpSetup}
\end{eqnarray}
where $\sigma_{{\rm{Exp}}}$ would correspond to the experimental data. Expanding right hand side to $\mathcal{O}(\alpha)$ we see
\begin{eqnarray}
\frac{d}{dt}\left(\frac{d \sigma_{{\rm{Exp}}}} {d \sigma^{\rm{SM}}_{\rm{pert}}}\right)  =1+2\Delta \alpha_{\text{had}} + O(\alpha^2) \, .
\label{eq:RunHadSM}
\end{eqnarray}
Implicit in the above expansion is that any physics not accounted for in $d \sigma^{\rm{SM}}_{{\rm{pert}}}$, but present in the data will be absorbed into the definition of \ah. It is therefore mandatory to calculate $d \sigma^{\rm{SM}}_{\rm{had}}$ as accurately as possible in order to minimize unwanted inclusion of known physics (for example inclusion of NNLO effects in $\alpha$ from perturbative physics). Specifically the ~\muone~literature frequently quotes an error target of 10 ppm $(10^{-5})$ on the cross section as the desired goal  for the theoretical accuracy \cite{Abbiendi2017,Abbiendi:2677471,Alacevich2019,Dainese:2019xrz,Fael:2019nsf}. 
This value is motivated by considering degradation of the fitting function by inclusion of a systematic uncertainty. Studies have been performed which suggest that including systematic shifts proportional to  LO, included as parameters in the fit, do not significantly degrade the results beyond the initial 0.3\% value~\cite{CernTalk}. This can be easily understood, since the essential change is to include a fourth parameter in the fitting model, however with 30 well measured points this does not lead to a significant decrease in the fitting ability. 

The aim of this paper is to study in greater detail the nature of the theoretical quantity which would be fit using the proposed experimental procedure above. In general there are two types of missing theoretical components to equation~\eqref{eq:RunHadSM}. Firstly, as indicated above there are missing higher order corrections in the SM itself. Secondly, in the case physics BSM exists, the theoretical expansion of the differential cross section could be modified at $\mathcal{O}(\alpha)$ or $\mathcal{O}(\alpha^2)$. We therefore capture all missing theoretical information in the following equation 
\begin{eqnarray}
\frac{d}{dt}\left(\frac{d \sigma_{{\rm{Exp}}}} {d \sigma^{\rm{SM}}_{\rm{pert}}}\right)  =1+2(\Delta \alpha_{\text{had}}+\Delta \alpha_{\text{HO}}+\Delta \alpha_{\text{BSM}}) , 
\label{eq:RunHadSMFull}
\end{eqnarray}
where $\Delta \alpha_{\text{HO}}$ defines all unsubtracted pieces of the SM (for instance, electron mass effects, N3LO, etc.) and $\Delta \alpha_{\text{BSM}}$ corresponds to a model dependent BSM correction.
The fit to experimental data will therefore simultaneously fit the target signal, \ah~ and the additional pieces. 
When integrated to obtain $a^{\rm{HLO}}$ the unsubtracted terms modify the result as follows
\begin{eqnarray}
a^{\text{HLO}} \rightarrow a^{\text{HLO}}  + \delta a^{\text{HLO}} \, , 
\label{eq:modhlo}
\end{eqnarray}
where $ \delta a^{\text{HLO}} $ captures the integrated pieces which do not arise from the hadronic running of $\alpha$ 
\begin{eqnarray}
\delta a^{\text{HLO}}&=&\delta a_{\text{HO}}^{\text{HLO}}+\delta a_{\text{BSM}}^{\text{HLO}}\, , \\
&=&\frac{\alpha}{\pi} \int_{0.3}^{0.932} dx (1-x) \left(\Delta \alpha_{\text{HO}}+\Delta \alpha_{\text{BSM}}\right) \, .
\label{eq:delHL0}
\end{eqnarray}
We have also specifically included the integration bounds of the \muone~ fiducial volume. The accuracy on the extraction of  $a^{\rm{HLO}}$ is thus intimately related to the size of $\delta a^{\text{HLO}}$, which should be compared to the 0.3\% (fitting) error target.  In the subsequent sections we will estimate the impact of $\delta a_{\text{HO}}^{\text{HLO}}$ using estimates of higher order calculations, and $\delta a_{\text{BSM}}^{\text{HLO}}$ for general tree-level and loop induced new physics scenarios. 

\section{Results}
\label{sec:results}

\subsection{Impact of missing higher order corrections} 

We begin by studying the impact of missing higher order terms in the SM. 
Formally these are well-defined by their inverse (since we know which terms of the SM are included).
 For practical definitions, until the completion of higher calculations becomes available,  one can only construct some approximate form of missing higher order corrections. 
The simplest function to define is a flat correction relative to LO, i.e. $\Delta \alpha^{\text{LO}}_{\text{HO}} = c $
and correspondingly 
\begin{eqnarray}
\delta a_{\text{LO}}^{\text{HLO}} = c  \frac{\alpha}{\pi} \int_{0.3}^{0.932}  (1-x) dx \, ,
\end{eqnarray}
This integral can be readily computed, yielding 
\begin{eqnarray}
\delta a_{\text{LO}}^{\text{HLO}}(c) = 0.243\, \frac{\alpha c}{\pi} \, .
\end{eqnarray}
The stated theoretical accuracy of 10 ppm would correspond to $c = 5 \times 10^{-6}$ (accounting for the factor of two in eq.~\eqref{eq:RunHadSMFull}), this results in $\delta a_{\text{LO}}^{\text{HLO}}(5 \times 10^{-6})  = 28 \times 10^{-10}$,  which is approximately equal to $\Delta a_{\mu}$ the current difference between theory and experiment for the $g-2$ measurement, and is far too large for a useful extraction of $a^{\text{HLO}}$.
In order to reduce the impact of the missing higher order corrections to the level of the current uncertainty on $a^{\text{HLO}}$ of $ \sim 2 \times 10^{-10} $ (adjusted from the full error for the fiducial volume of \muone), $c$ would have to be $3.5 \times 10^{-7}$, corresponding to a theoretical accuracy of 0.7 ppm on the differential cross section. 
In principal missing higher orders of this form (proportional to LO) enter the fit as a constant term, and therefore generate a non zero value of $c_0$ in the expansion in $t$. These terms could therefore be treated as a theoretical systematic uncertainty in much the same way as other systematic uncertainties are handled. 

A more worrisome class of corrections are missing terms with a dynamic $t$ dependence across the fiducial volume of the \muone~ experiment which cannot simply be absorbed into a constant term in the fit. In order to investigate the potential impact of these types of terms we estimate the size of various higher order corrections by computing the leptonic running of $\alpha$ raised to the appropriate power 
\begin{align}
\Delta \alpha_{i,\text{HO}}^{\text{approx}}(t) &=  \kappa_i \Delta \alpha_{{\rm{lep}}}^{i}(t)  \, ,
\label{eq:daaprox}
\end{align}
where $\Delta \alpha_{{\rm{lep}}}^{i}(t)$ is defined in Appendix~A.
We note that these pieces correspond to a single diagram (rescaled by LO) from the $i^{\rm{th}}$ order correction in the perturbative series (namely the equivalent topology of Fig.~\ref{fig:alph} with $i$ bubble insertions along the photon line). While this diagram gives an idea of the order of magnitude of a missing $i^{\text{th}}$ order correction, the full result could be smaller (due to cancellations with other diagrams) or larger. Therefore we vary our estimate over some set range, by multiplying it with a factor of $\kappa_i$. A reasonable range for $\kappa_i$ can be best determined upon  completion of the NNLO computation, but for now we will take $\kappa_i$ in the range $\{ 1/5- 5\}$. We can validate our estimate at NLO $(i=1)$, by comparing it to the available NLO calculation. We find that the exact value $\delta a_{1,\text{HO}}^{\text{exact}}=1.4 \times 10^{-6}$ lies well within our estimated range $\delta a_{1,\text{HO}}^{\text{approx}}=\left(0.6-16\right) \times 10^{-6}$, where the former was obtained by interpolating the data points presented in~\cite{Alacevich2019} and adjusting the lower integration bound to be $x=0.373$, due to the restricted range of the NLO data. 

Using this approximate form, we compute $\delta a_{i,\text{HO}}^{\text{approx}}$ using eq.~\eqref{eq:daaprox} in eq.~\eqref{eq:delHL0}, as a result we estimate that the unknown higher order contributions to $\delta a^{\text{HLO}}$ have the following sizes:  
\begin{align}
|\delta a_{i=2,\text{HO}}^{\text{approx}}|&=\left(0.5-13 \right) \times 10^{-8} \, , \\
|\delta a_{i=3,\text{HO}}^{\text{approx}}|&=\left( 0.4-9 \right) \times 10^{-10} \, , \\
|\delta a_{i=4,\text{HO}}^{\text{approx}}|&=\left(0.3-6 \right) \times 10^{-12} \, .
\end{align}
Our estimates corroborate current beliefs that an NNLO ($i=2$) calculation is essential for the success of the experiment. 
However, the impact on $a^{\text{HLO}}$ from the $i=2$ contribution is estimated to be 
around 100 times the target uncertainty of $2\times 10^{-10}$. It is not until $i=3$, (corresponding to an approximate N3LO calculation) that the estimated corrections are around the desired accuracy of $2\times 10^{-10}$. These statements are in line with those outlined in recent \muone~ Letter of Intent (LOI), Ref~\cite{Abbiendi:2677471}, which (using the 10 ppm target) determined the need for both NNLO and resummation effects to achieve the theoretical error target. Both the current literature and our analysis above suggest that precision significantly beyond NNLO will be required. Before attempting to quantify what terms are needed we first determine the impact of the higher order estimate on the differential distribution in terms of ppm accuracy on the cross section. 

\begin{center}
\begin{figure}
\includegraphics[width=8cm]{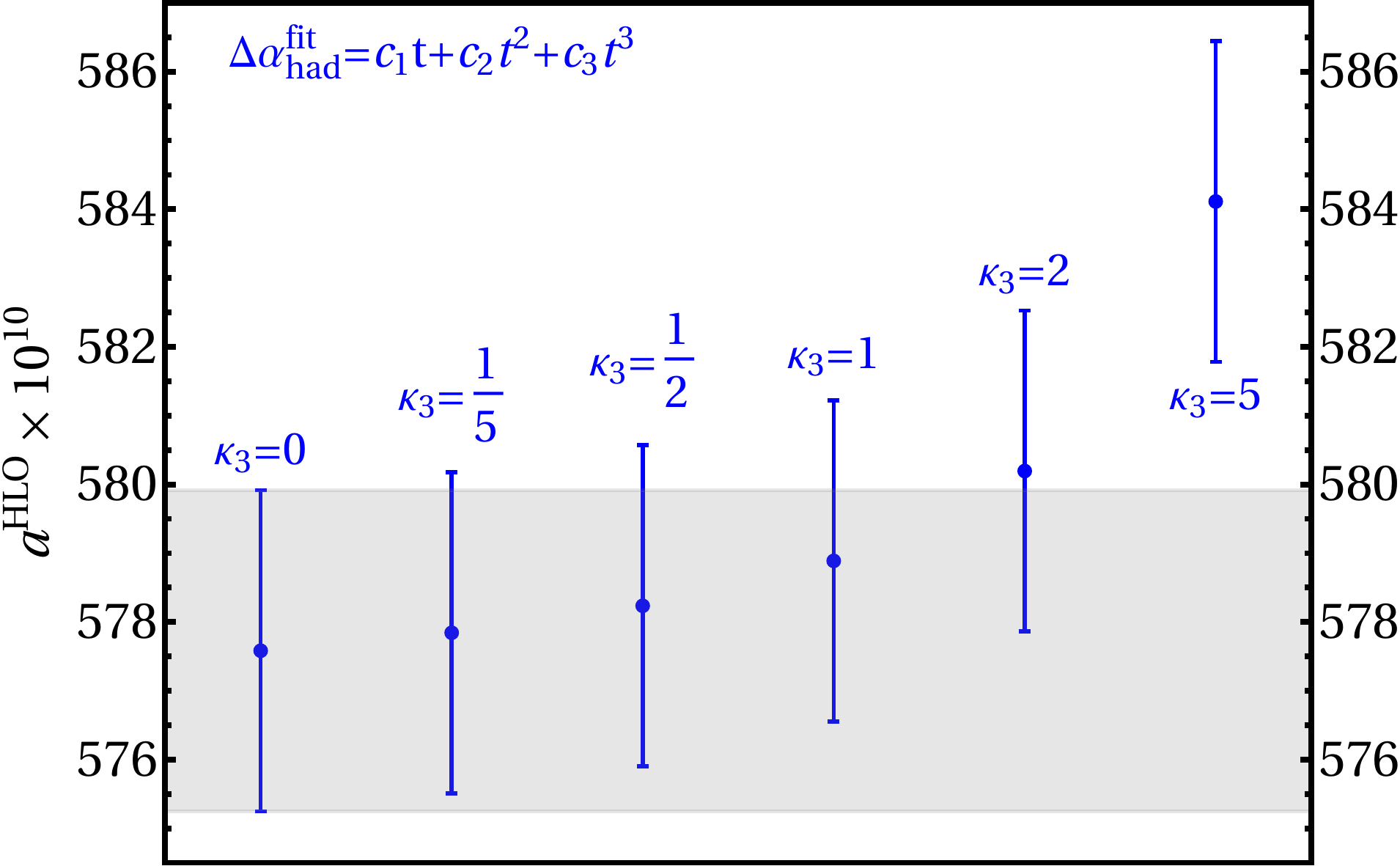}
\caption{Different values of the leading hadronic contribution, depending on the size of $\kappa_3$ multiplying the approximate  N3LO correction, defined according to eq. \eqref{eq:daaprox}.}
\label{fig:amuPlot}
\end{figure}
\end{center}

To do so we perform the following analysis, we generate data for \ah~ using {\tt{hadr5n12}} and for $\Delta \alpha_{i,\text{HO}}^{\text{approx}}$ with a given value of $\kappa_i$. We then proceed to fit the sum of these two terms together with a flat statistical uncertainty of 10 ppm using a third order polynomial fit 
\begin{eqnarray}
\Delta \alpha_{\text{had}}^{\text{fit}} = \Delta \alpha_{\text{had}} +  \kappa_i (\Delta \alpha_{{\rm{lep}}}^{i})^3  = \sum_{i=1}^{3} c_i t^i \, ,
\end{eqnarray}
which is then integrated through eq.~\eqref{eq:HL0}. We obtain fitting uncertainties of $2.3\times10^{-10}$, which matches those found in previous \muone~ studies.  Our results obtained for various values of $\kappa_3$ are shown in Fig.~\ref{fig:amuPlot}, which included the pure hadronic contribution ($\kappa_3=0$). We observe that the errors arising from the fit do not depend on $\kappa_3$ but that the central value of $a^{\rm{HLO}}$ can be significantly shifted. Especially if $\kappa_3 \geq 2$ the shift exceeds the fitting error of the purely hadronic data with $\kappa_3=0$ and therefore is the dominant theoretical uncertainty. For $\kappa_3 = 1.5$ the shift in $a^{\text{HLO}}$ is 2$ \times 10^{-10}$ and we find that the mean value of $\Delta \alpha_{3,\text{HO}}^{\text{approx}}$ is $0.62 \times 10^{-6}$ and the maximum value is 1.5 $\times 10^{-6}$. This indicates that a function which contributes at the level of around 1.2 ppm is sufficient to induce a change in $a^{\text{HLO}}$ greater than $2\times 10^{-10}$. 

The two estimates presented thus far in this section suggest that a precision of around 1 ppm will be required in order to ensure that the extracted value of $a^{\text{HLO}}$ is not altered by the presence of missing higher order corrections at the level of $2\times 10^{-10}$ or greater. Further, our second approximate form suggested this precision occurred around the N3LO level.
Since an N3LO calculation remains a daunting task, it is natural to investigate whether a suitable approximation could be constructed to capture the dominant impact of this term in the perturbative expansion. 
Such approximate forms were discussed in the LOI~\cite{Abbiendi:2677471} and  compared to the 10 ppm standard. Here we reinvestigate the issue, in light of the results of the previous section. 
In order to do so we decompose the perturbative expansion in $\alpha$ to order $n$ as follows 
\begin{eqnarray}
\sigma^{(n)} =  \sigma^{(0)}  \sum_{m = 0}^{n}\left[ \sum_{i=0}^{m}\sum_{j=0}^{m} \kappa_{i,j}^{m}(t) \left(\frac{\alpha}{\pi}\right)^m  L^i \ell^j \right] \, ,
\label{eq:sigb}
\end{eqnarray}
where we have used the notation of Ref.~\cite{Abbiendi:2677471}, parameterizing the cross section in terms of two IR logarithms $L= \log(-t/m^2)$ and $\ell=-2\log{(2\Delta\omega/s)}$. $\kappa_{i,j}^{m}(t)$ defines the coefficient of the logarithms, which is in general a function of the kinematic variables (of which we are primarily concerned with the $t$ dependence).  $\Delta \omega$ is related to the experimental definition of photon/leptons and its discussion is beyond the scope of this paper. 
Steps outlining the resummation of these logarithms are discussed in  Ref.~\cite{Abbiendi:2677471}. Rather we take as a starting point the
stated theoretical uncertainty from Ref.~\cite{Abbiendi:2677471} after the proposed resummation techniques have been applied, which is given by the term 
\begin{eqnarray}
\kappa_{2,0}^{3} \left(\frac{\alpha}{\pi}\right)^3  L^2 \, ,
\label{eq:muoneer}
\end{eqnarray}
i.e. $L^2$ with no enhancement by a $\ell$ power and with $\kappa_{2,0}^{3}$ of $\mathcal{O}(1)$. Our aim is thus to relate the uncertainty induced by this term to the findings of our previous analyses. 
It is straightforward to  relate this expansion to our previous estimate of the missing higher order corrections. Our estimate was constructed from one-loop bubble integrals, which contain logarithms of the form 
\begin{eqnarray}
L' = \log{\left(-\frac{1+\sqrt{1-t/(4m_e^2)}}{1-\sqrt{1-t/(4m_e^2)}}\right)} \, ,
\end{eqnarray}
(see Appendix A for the full form). Over the range of phase space available to \muone~ one can write to a good approximation 
\begin{eqnarray}
L' \approx L = \log{\left(-\frac{t}{m_{e}^2}\right)} = \log{\left(\frac{x^2}{1-x}\frac{m_{\mu}^2}{m_{e}^2}\right)} \sim 10 \, .
\end{eqnarray}
Hence, our previous estimation was of the form $a(t) L^3 + b(t) L^2 + c(t) L + d(t)$, which we note contains cubic powers of $L$, and is therefore of higher order than eq.~\eqref{eq:muoneer}.
Motivated by our previous study, and the error estimate of Ref.~\cite{Abbiendi:2677471} we study the functions
\begin{eqnarray}
\Delta \alpha_{j,n,\text{HO}}^{{L}}= \frac{1}{2}\left(\frac{\alpha}{\pi}\right)^n  \kappa^{n}_{j,0} L^j \, ,
\label{eq:logb}
\end{eqnarray}
where $0 \le j \le n$, and for simplicity we take $\kappa$ to be an unknown constant. The factor of $1/2$ is inserted to ensure a consistent definition of $\Delta \alpha_{\text{HO}}$ in eqs.~\eqref{eq:modhlo} and ~\eqref{eq:sigb}, which allows us to quickly relate $\kappa$ to the expansion coefficient of a particular term in the cross section. 
This then resembles our previous estimate in terms of the logarithmic structure (with all rational functions of $t$ dropped), and the stated error estimate from Ref.~\cite{Abbiendi:2677471} which starts at $j=2$ (for $n=3)$.

\begin{center}
\begin{table}
\begin{tabular}{|c|c|c|c|c|}
\hline
  &   $(\alpha/\pi)^n L^n$ &    $(\alpha/\pi)^n L^{n-1}$ &   $(\alpha/\pi)^n L^{n-2} $ &  $(\alpha/\pi)^n L^{n-3}$  \\
\hline
 $n=2$   & $980$ & $88$ & $8$ & $-$   \\
\hline
 $n=3$ & $26$ & $2.3$ & $0.21$ & $0.019$   \\ 
\hline
 $n=4$ & $0.67$ & $0.059$ & $0.0053$ & $0.00048$   \\ 
\hline
\end{tabular}
\caption{$|\delta a| \times 10^{10}$ values for various powers of $\alpha$ and $L$ with $\kappa_{a,b}^{n} =1$ for all contributions.}
\label{tab:aHOI}
\end{table}
\end{center}

We perform the same analysis as the previous estimate, namely performing a combined fit to \ah~ and $\Delta \alpha_{j,n,\text{HO}}^{{L}}$ and integrating the total to obtain a modified $a^{\rm{HLO}}$. 
We present the difference from the \ah~only result (in units of 10$^{-10}$) in Table~\ref{tab:aHOI} where we have set $ \kappa^{n}_{j,0} = 1$ for simplicity. We present results for $n=2,3,4$ and $0\le j \le n$. 
Of particular interest is the comparisons between our previous third order estimate, and that obtained here. We see that the shift induced by the $\alpha^3 L^3$ term is significantly larger than the third order approximate constructed from the leptonic running ($\sim$ 25 compared to 1). This is primarily since the leptonic running scales like $1/3$ compared to $\Delta \alpha_{3,3,\text{HO}}^{{L}}$, when raised to the third power this causes a suppression of 27. 
In the leptonic running there is also a partial cancellation between the $L^3$ and $L^2$ terms, which suppresses the contribution by roughly a factor of two. This partial cancellation is mimicked by the factor of 1/2 in eq.~\eqref{eq:logb}.

We observe that the $\alpha^3 L^2$ term (with a coefficient of 1) contributes around $2 \times 10^{-10}$ to $\delta a$. This strongly suggests that this term will be required in order to achieve the desired accuracy of \muone~ (unless for some reason the coefficient was significantly smaller than 1). 
The $\alpha^3 L$ term contributes around $0.2 \times 10^{-10}$ to $\delta a$ and is therefore extremely sensitive to the true value of $\kappa$ as to whether or not is contribution is mandated. 
The coefficient of 
$\alpha^4 L^3$ is of a similar (albeit smaller) size its presumed importance (or lack thereof) should be easier to quantify once more is known about the perturbative expansion.

\begin{center}
\begin{table}
\begin{tabular}{|c|c|c|c|c|}
\hline
 &   $(\alpha/\pi)^n L^n$ &    $(\alpha/\pi)^n L^{n-1}$ &   $(\alpha/\pi)^n L^{n-2} $ &  $(\alpha/\pi)^n L^{n-3}$  \\
\hline
$n=2$   & $0.002$ & $0.02$ & $0.3$ & $-$   \\
\hline
 $n=3$ & $0.08$ & $0.9$ & $10$ & $100$   \\ 
\hline
  $n=4$ & $3$ & $30$ & $400$ & $4000$   \\ 
\hline
\end{tabular}
\caption{$\kappa$ values for individual coefficient at which $|\delta a|$ exceeds the desired accuracy of $2 \times 10^{-10}$.}
\label{tab:kappa}
\end{table}
\end{center}
In order to present this information in a more usable format (after completion of future higher orders) we compute the value of $\kappa$ required for each term (taken individually) to be sufficient to alter the fitted $a^{\rm{HLO}}$ by greater than or equal to $2\times 10^{-10}$. 
These values of $\kappa$ are summarized in Table~\ref{tab:kappa}. For example, we see that if the $\alpha^3 L$ term has a coefficient greater than $10$ it will need to be included in the theoretical calculation. 
If the coefficients are of $\mathcal{O}(1-10)$ then $\alpha^3 L^2$, and  $\alpha^3 L$, should be included. Upon understanding of the perturbative structure at $n=2$ and $n=3$  it may be possible to predict more accurately whether the $\alpha^4 L^3$ terms are needed. Presumably, if the technology exists to determine the third order up to single logarithmic accuracy, similar techniques may be utilized to determine the $\alpha^4 L^3$ term. We recall that in reality the coefficients of these sub-leading logarithms are themselves functions of the external kinematics, and therefore modeling them as a constant is somewhat risky. Needless to say, once more is known about the lower order terms in the expansion it will be easier to make more predictive comments about exactly which coefficients are needed. In summary it seems that the knowledge of $\alpha^3 L^2$ is mandatory, and that the $\alpha^3 L$ and $\alpha^4 L^3$ terms need a more robust argument as to judge the maximum size of their coefficient in the full perturbation theory  (ideally with an actual calculation). 

As a final issue we comment on the role of fitting in determining our results above. We recall that the numbers computed in this section were obtained by integrating $\Delta \alpha_{\rm{had}}^{\rm{fit}}$ which corresponded to the combination of \ah~and the chosen higher order correction with an error given by $5\times 10^{-6}$ in the fit. It is interesting to compute the un-fitted corrections to $\delta a$ arising from eq.~\eqref{eq:logb} integrated term by term, i.e.
\begin{eqnarray}
\delta a^{n}_{i} = \frac{1}{2}\left(\frac{\alpha}{\pi}\right)^{n+1} \int_{0.3}^{0.932} (1-x)  \log{\left(\frac{x^2}{1-x}\frac{m_{\mu}^2}{m_{e}^2}\right)}^i  dx  \nonumber\\
\end{eqnarray}
where we have set $\kappa_{i,0}^{n}=1$.  Focusing on $n=3$ and $i=2,1$ we find 
\begin{eqnarray}
\delta a^{3}_{2} = 3.7 \times 10^{-10}, \quad  \delta a^{3}_{1} = 0.36 \times 10^{-10}
\end{eqnarray}
which should be compared to $2.3$ and 0.21 $\times 10^{-10}$ (Table~\ref{tab:aHOI}) for the respective fitted values. We observe that fitting the logarithms onto a cubic polynomial reduces the ``pure" impact of the pieces by around a factor of two. Other fitting functions have been investigated for \ah~ (and are beyond the scope of this work), but it would be an interesting study to investigate if these terms could be further suppressed by modifications to the fitting procedure.

We present a summary of our various findings in Table~\ref{tab:ppm}. The three different types of functions all paint a broadly similar picture. That is, in order to extract $a^{\rm{HLO}}$ with a systematic uncertainty 
from missing higher order corrections less than or equal to $2\times 10^{-10}$ requires control of the differential $t$-distribution at around the 1 ppm level. We classified the size of the coefficients needed in a perturbative expansion of $\alpha$ and $\log{(-t/m_e^2)}$. Anticipating the size of the coefficients at around $\mathcal{O}(1-10)$ mandates terms up to order $\alpha^3 L$ and possibly $\alpha^4 L^3$. As more theoretical work is completed it will be possible to determine the likely size of missing coefficients more accurately. Finally, we can estimate the theoretical systematic uncertainty arising from a function which has a mean value corresponding to the 10 ppm target (as originally proposed~\cite{Abbiendi:2677471}), 
performing the same analysis as above (using the $\alpha^3 L^2$ template function) results in a shift of $\delta a_{\rm{HO}}^{\rm{HLO}} =16\times 10^{-10}$, with other choices of templates functions resulting in similar values.  
 
\begin{center}
\begin{table}
\begin{tabular}{|c|c|c|c|}
\hline
Function  & Mean $[\times 10^{-6}]$ & Maximum $[\times 10^{-6}]$ &$\delta \sigma/\sigma^{(0)}$ [ppm]   \\
\hline
$\Delta \alpha_{3,2,\rm{HO}}^{L} [\kappa = 0.9]$   & $0.67$ & $1$ & 1.3 \\
\hline
$\Delta \alpha_{3,1,\rm{HO}}^{L} [\kappa = 10]$  & $0.67$ & $0.83$& 1.3 \\
\hline
$\Delta \alpha_{4,4,\rm{HO}}^{L}[\kappa =3 ]$ & $0.66$ & $1.4$& 1.3 \\
\hline
$\Delta \alpha_{\rm{HO}}^{\rm{LO}} $  & $0.35$ & $0.35$& 0.7 \\
\hline
$1.5 \times \Delta \alpha_{\rm{lep}}^3(t)$   & $0.62$ & $1.47$& 1.2 \\
\hline
\end{tabular}
\caption{Average and maximal sizes of different contributions in ppm $(10^{-6})$, which induce a theoretical systematic uncertainty that is larger then the targeted accuracy of $2 \times 10^{-10}$.}
\label{tab:ppm}
\end{table}
\end{center}

\subsection{Tree-level BSM Contributions}

The previous section discussed the precision needed in the SM to enable an accurate extraction of ~\ah. There is a second component to eq.~\eqref{eq:RunHadSMFull} arising from putative BSM physics. In this section we analyze the potential impact of different types of models.  We begin by discussing BSM contributions which may enter first at tree-level. A tree-level exchange connects the two lepton lines, and therefore corresponds to an example like the rightmost diagram in Fig.~\ref{fig:alphBSM}. In order to have avoided detection the new boson exchanged between the leptons must either be very heavy or have suppressed couplings to SM matter.  The simplest examples correspond to the exchange of either a spin-0 scalar or spin-1 gauge boson. 
For the case of a scalar (or pseudo-scalar) the couplings scale with lepton masses. This assumes
minimal flavor violation, in which case the only flavor violating spurions are the Yukawa matrices. Given the smallness of the electron Yukawa we focus instead on the exchange of a spin-1 gauge boson. 
In order to study a (simple) relevant example we consider a vector gauge boson arising from an 
an additional $U(1)_X$ symmetry (referred to as a Secluded $U(1)$) that mixes with the standard model photon (a dark photon).
Originally such models were proposed as good candidates to explain the difference in $g-2$~\cite{Pospelov:2008zw}, however much of the desired parameter space is now 
excluded by other measurements~\cite{Bauer2018}. However there is still an unconstrained region of parameter space in which the 
models could have some impact on $g-2$. 

A secluded $U(1)$ Lagrangian includes a new gauge boson $X_{\mu}$ which couples to the hypercharge gauge 
boson of the SM~\cite{Pospelov:2008zw,Bauer2018} 
\begin{eqnarray}
\mathcal{L} &=&- \frac{1}{4}\hat{F}_{\mu\nu}\hat{F}^{\mu\nu}-\frac{\epsilon'}{2}\hat{F}_{\mu\nu}\hat{X}_{\mu\nu}
-\frac{1}{4}\hat{X}_{\mu\nu}\hat{X}^{\mu\nu}\nonumber\\&&-g'y_{\mu}^{Y}\hat{B}^{\mu}+\frac{1}{2}\hat{M}_X^2\hat{X}_{\mu}X^{\mu} \, ,
\end{eqnarray} 
where $\hat{B}_{\mu}$ defines the hypercharge gauge boson, with corresponding field strength tensor
$\hat{F}_{\mu\nu}$, gauge coupling $g'$ and the hypercharge current $j_{\mu}^{Y}$ defines the coupling 
to SM fermions. The hats on the fields indicate that the relevant fields are not canonically normalized and 
require rotation to the mass basis. This results in a redefinition of the SM $Z$ boson mass, a massless photon 
and a massive dark photon (denoted by $A'$). The dark photon couples universally to all SM quark and lepton flavors with a suppression 
given by $\epsilon = \epsilon' \cos\theta_w$. 

Recalling Eq.~\eqref{eq:RunHadSMFull} we see that unaccounted BSM physics would be absorbed into the hadronic running of $\alpha$. For tree-level BSM physics we find a contribution of the form
 
\begin{eqnarray}
\Delta \alpha_{\text{BSM}}= \frac{\Re[M_{\text{QED}}M_{\text{BSM}}]}{\vert M_{\text{QED}} \vert^2},
\end{eqnarray}
which corresponds to the interference term between the SM and BSM contributions. We neglect the BSM squared term, assuming that it remains small. 
\begin{widetext}
\begin{center}
\begin{figure}
\includegraphics[width=15cm]{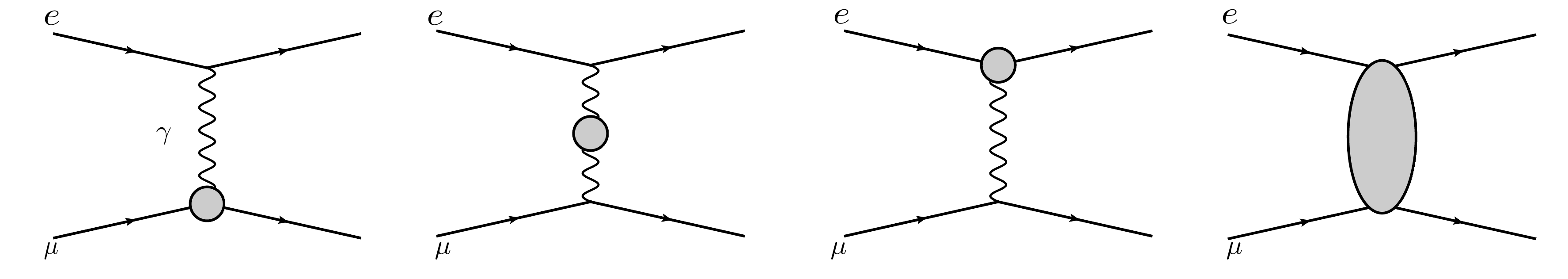}
\caption{Topologies illustrating the possible insertion of BSM physics. }
\label{fig:alphBSM}
\end{figure}
\end{center}
\end{widetext}

For the dark photon model described above the LO interference is simple to compute, indeed since the only modifications are to the $t-$channel propagator and the coupling, the ratio is given by 
\begin{eqnarray}
\Delta \alpha_{\text{DP}} = \frac{\epsilon^2 t}{t-m_{A'}^2} \, ,
  \label{eq:DPtree-level}
\end{eqnarray}
where $m_{A'}$ is the mass of the new boson. We note that  formally the width of the dark photon, $\Gamma_{A'}$ enters into the above expression. The width, however, is suppressed by an additional factor of $\epsilon^2$. As a result the effects of the width enter at the same level as the BSM contribution squared ($\epsilon^4$) which we neglect consistently. 

\begin{figure}[t]
 \centering
  \includegraphics[scale=0.5]{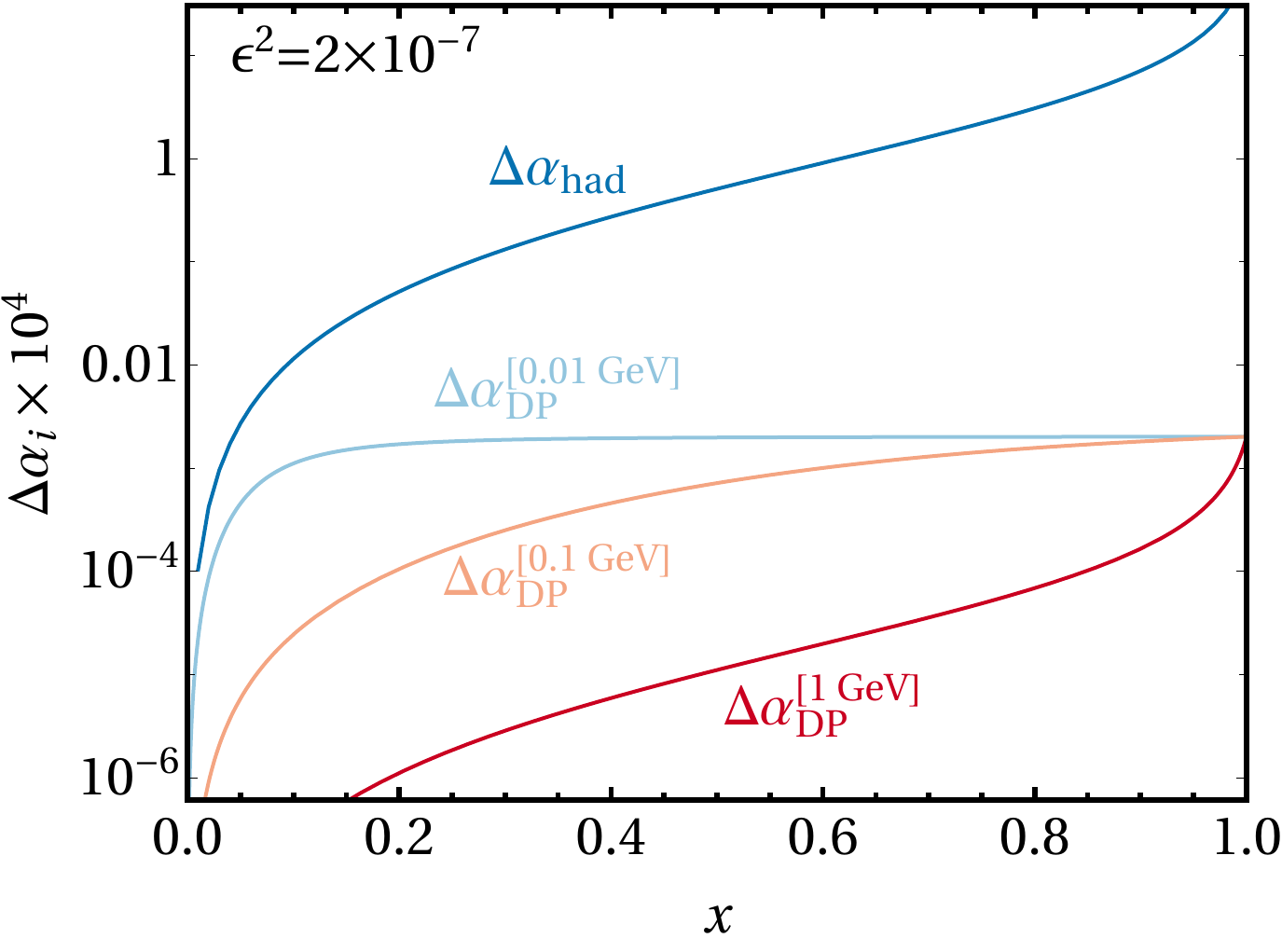}
 \caption{The plot shows the Leading Order hadronic contributions \ah~ in comparison with various dark photon models $\Delta \alpha_{\rm{DP}}^{[m_{A'}]}$.}
 \label{fig:DPplot}
\end{figure}

Our results are presented in Fig.~\ref{fig:DPplot} where we present $\Delta \alpha^{[m_{A'}]}_{\text{DP}}$ as a function of $x$. We have chosen 
$\epsilon^2=2\times 10^{-7}$ which is close to the current exclusion limits for the Secluded $U(1)_X$ model~\cite{Bauer2018} and present curves for three different mass choices, $m_{A'} = \{0.01 ,0.1, 1\}$ GeV. 
It is clear that both the shape, and relative importance compared to \ah~is strongly dependent on $m_{A'}$. Since in the spacelike region $t$ is negative, the denominator of eq.~\eqref{eq:DPtree-level} never diverges and therefore increasing the mass lowers the overall impact. As $|t| > m_A$ the ratio approaches an asymptotic value of $\epsilon^2$.

\begin{table}
\begin{tabular}{|c|c|c|c|}
\hline
$\epsilon^2$ &  \multicolumn{3}{|c|}{$2 \times 10^{-7}$} \\
\hline
$m_{A'}[GeV]$  & $0.01$& $0.1$ & $1$ \\
\hline
$\delta a_{BSM}^{\text{HLO}}$ & $1.1 \times 10^{-10}$ & $4.6 \times 10^{-11}$ & $1.3 \times 10^{-12}$  \\
\hline 
\end{tabular}
\caption{Integrated contributions to $\delta a_{BSM}^{\text{HLO}}$ stemming from three different dark photon models.}
\label{tab:aDP}
\end{table}

We present results for the integrated contribution to $\delta a_{BSM}^{\text{HLO}}$ in Table~\ref{tab:aDP}. Assuming that the \muone~experiment can reach their expected sensitivity of $2\times 10^{-10}$, we see that the lightest dark photon model contributes at the level of this uncertainty. Since this relevant parameter space is at the edge of the current exclusion limits~\cite{Bauer2018}, future dark photon searches will have likely excluded all relevant parameter space by the time the \muone~experiment is performed. 

In contrast if the BSM sector is expanded to include BSM matter content e.g. by including a dark matter candidate~\cite{TuckerSmith:2001hy,Izaguirre:2015zva}, the incorporated potential decays to light dark sector fermions (or scalars) can alter the width of the dark photon. As a result the decays to SM matter can be suppressed, causing an overall lowering of the ability of direct searches to constrain 
the dark photon parameter space. Such an extended dark sector was recently proposed as a method of explaining $g-2$ while evading some existing searches~\cite{Mohlabeng2019}. 
Since the \muone~experiment is only marginally dependent on the decay width of the dark photon such extended BSM models could make sizable contributions, while evading bounds from direct searches. Consequently a careful examination of the current dark photon bounds will be necessary, once the \muone~data is analyzed.     
 
Finally we note that models in which the BSM mediating particle is heavy can also be interpreted using Eq.~\eqref{eq:DPtree-level} with the replacement $\epsilon^2 \rightarrow \mathcal{O}(1)$ for EW scale couplings.  In this instance we see the suppression is given by $t/(t-m^2)$. It is thus clear that heavy new physics will effectively look like $t/m^2$ in the fiducial region of the \muone~ experiment. LEP limits~\cite{Ackerstaff:1996sf,Acciarri:2000uh,Abbiendi:2003dh,Abdallah:2005ph,Schael:2006wu} on contact interactions already impose $m \gtrsim \mathcal{O}$ TeV. Which is sufficient to suppress heavy new physics to such a level as to be neglected in the \muone~analysis.

\subsection{Loop-level BSM Contributions}

We now consider the possibility that the SM prediction is modified at the one-loop level. This changes eq.~\eqref{eq:RunHadSM} to the following
\begin{eqnarray}
\frac{d}{dt}\left(\frac{d \sigma_{{\rm{Exp}}}} {d \sigma^{\rm{SM}}_{\rm{pert}}}\right) & =&1+2\Delta \alpha_{\text{had}} 
+2\alpha \frac{d}{dt}\left(\frac{d \sigma^{\rm{BSM}}}{d \sigma^{\rm{LO}}}\right) \nonumber\\&&+ O(\alpha^2) \, ,
\label{eq:bsmexp1l}
\end{eqnarray}
where $d\sigma^{\rm{BSM}}$ defines the BSM physics contribution, we note that, due to the implicit insertion of a factor of $\alpha$ in the definition of \ah~, both terms in eq.~\eqref{eq:bsmexp1l} are of the same (formal) order in the perturbative expansion of the ratio. 
As illustrated in Fig.~\ref{fig:alphBSM} there are four possible insertions of BSM interactions in the basic LO topology. They can be included on either of the individual lepton lines, the photon propagator, or by connecting the two lines together. We can thus expand $d\sigma^{\rm{BSM}}$ as follows
\begin{eqnarray}
d \sigma^{\rm{BSM}} = \left(\Delta \alpha^{\gamma}_{BSM}  + \sum_{\ell=e,\mu} \Delta \alpha^{\ell}_{BSM}  + \Delta \alpha^{e,\mu}_{BSM} \right) d\sigma^{\rm{LO}} \nonumber \, ,
\end{eqnarray}
where $\Delta \alpha^{i}_{BSM}$ corresponds to the correction associated with particle content $i$ factored onto the LO differential cross section $d\sigma^{\rm{LO}}$. 
In this work we set $\Delta \alpha^{e,\mu}_{BSM} =0$, that is we neglect  box-type contributions in which the new physics connects the two lepton lines. 
This is primarily because they either represent a QED correction to a tree-level contribution (as discussed in the previous section) or involve couplings that are heavily constrained by lepton flavor violation, or are suppressed by the smallness of the electron Yukawa coupling (for exchange of scalar particles). Contributions from heavy new physics which are not flavor suppressed effectively reduce to a four-fermion contact interaction, which falls under the discussion of the previous section. Before detailing the calculation in specific models (the MSSM and a gauged $U(1)_{L_{\mu}-L_{\tau}}$ model) in the next section, we first outline the characteristics of the remaining contributions.

The contributions from one-loop BSM corrections to the photon propagator can be written as
\begin{eqnarray}
\Delta \alpha^{\gamma}_{BSM}= \Re[\Sigma^r(t)] \, ,
\label{eq:SelfEnergy}
\end{eqnarray}
where $\Sigma^r(t)$ defines the renormalized photon self energy, which in the on-shell scheme is given by $\Sigma^r(t)=\Sigma(t)-\Sigma(0)$. The equation above was obtained by expanding the photon propagator in $\alpha$
\begin{eqnarray}
D^{\mu \nu}=-i \frac{g^{\mu \nu}}{t} \left(1+\Sigma(t) + \mathcal{O}[\alpha^2] \right) \, ,
\label{eq:SEFF}
\end{eqnarray}
and introducing the form factor $\Sigma(t)$ as 
\begin{eqnarray}
\Sigma^{\mu \nu}(t)=i \left(t \, g^{\mu \nu} - (p_2-p_3)^\mu (p_2-p_3)^\nu \right) \Sigma(t) \, .
\end{eqnarray}
In addition we introduced the outgoing four-momenta of the electron and muon as  $(p_2-p_3)^2 =q^2=t$. In general heavy new physics will act much like the top-quark contribution to the 
self energy. For $t \ll m_t^2$, which corresponds to the majority of the range in $x$ (and includes the fiducial detector volume), these terms scale like $t/m_t^2$, so in general we do not expect significant contribution from heavy BSM physics from self-energy type corrections.

\begin{figure}[t]
  \centering
    \includegraphics[scale=0.35]{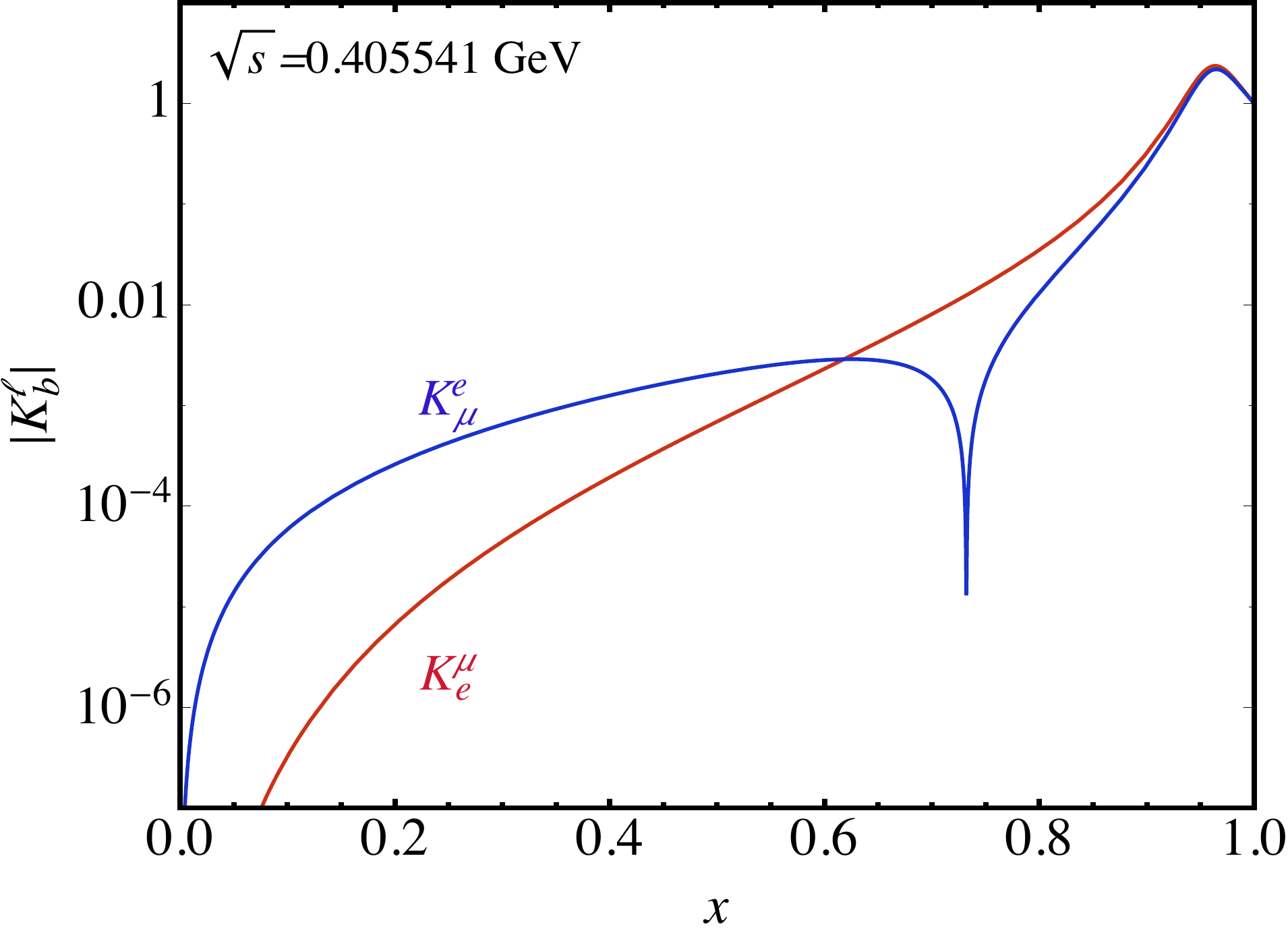}

  \caption{The kinematic factor $K$, evaluated for electron and muon vertices for the kinematics of the \muone~experimental setup.}
  \label{fig:kinfac}
\end{figure}

The corrections to the lepton-photon vertex lead to contributions of the form
\begin{eqnarray}
\Delta \alpha^{\ell}_{BSM}= F^{r,\ell}_{e}(t)+K^{\ell}_{b} F^{r,\ell}_{m}(t) \, .
\label{eq:Vertex}
\end{eqnarray}
The projection onto the tree-level matrix is obtained by first computing the unrenormalized electric and magnetic form factor in terms of its constituent Lorentz structures 
\begin{eqnarray}
\Gamma^\mu_{\ell}=-i \, e \left( \gamma^\mu F^{\ell}_{e}(t) + \frac{i \sigma^{\mu \nu} q_\nu }{2 m_l} F^{\ell}_{m}(t)  \right) \, ,
\label{eq:VFF}
\end{eqnarray}
with $\sigma^{\mu \nu}=i/2 \left[\gamma^\mu,\gamma^\nu \right]$. The renormalization at one-loop in the on-shell scheme is given by $F^{r,\ell}_e(t)=F^{\ell}_e(t)-F^{\ell}_e(0)$  and  $F^{r,\ell}_{m}(t)=F^\ell_{m}(t)$. We recall that  
$g-2$ for a lepton is defined as the magnetic form factor at zero-momentum $F^{r,\ell}_{m}(0)=a_{l}$. 
We can therefore quantify the impact on \ah~ from a BSM theory which contributes $a_{\mu}^{BSM}$ to $g-2$ using the following order of magnitude estimate 
\begin{eqnarray}
\Delta \alpha^{\ell}_{BSM} \sim K^{\ell}_{b} a_{\mu}^{BSM} \, ,
\end{eqnarray}
where we suppress the electric contribution for the following discussion. 
It is apparent that,  depending on the size of $K$  BSM physics may induce a change in $a^{\rm{HLO}}$ comparable to the BSM contribution to $g-2$. The kinematic factor $K$ thus plays a critical role in determining the overall impact of the BSM physics contribution to the interpretation of $a^{\rm{HLO}}$. It can be computed by taking the interference of the magnetic part of the form factor (along the fermion line $\ell$) with the rest of the amplitude defined as in the SM, with the pure SM amplitude and is 
\begin{eqnarray}
K^{\ell}_b=\frac{(2m_b^2+t)t}{2(m_e^2+m_\mu^2-s)^2+2st+t^2} \, ,
\end{eqnarray}
where $b$ is the mass of the ``spectator fermion", which couples via the SM vertex in the diagram. 
We plot the magnitude of the kinematic factor in Fig.~\ref{fig:kinfac} for both muon and electron vertices. It is clear that the kinematic factor, for the center of mass energy proposed by the  \muone~ collaboration, $\sqrt{s}=0.4055541$ GeV, results in a significant suppression of the magnetic part of the form factor over the majority of the available phase space. Therefore we can see that  $\Delta \alpha^{\ell}_{BSM} \ll  a^{BSM}$ and in general a model which seeks to explain $g-2$ by introducing a loop-level contribution of the form  $a^{BSM} \sim 20 \times 10^{-10}$ will contribute a negligible amount to the extraction of \ah~ in $\mu e$ scattering.

The argument above assumes that both $F^{r,\ell}_{e}(t)$ and $F^{r,\ell}_{m}(t)$ do not become sufficiently large at any point in the \muone~phase space regime probed. In order to demonstrate the suppression in a realistic setting we provide an example calculation in the MSSM and a gauged $U(1)_{L_{\mu}-L_{\tau}}$ model in the next section.

\section{MSSM}
\label{sec:MSSM}

Supersymmetric (SUSY) theories offer a compelling UV complete extension to the SM, of which the Minimally Supersymmetric Standard Model (MSSM) provides a relatively simple realization of SUSY by introducing a limited number of new parameters~\cite{Martin1997}. In general, while the MSSM is by now severely constrained by the LHC, it is rather easy in this model to introduce new contributions to the muon anomalous magnetic moment, which can be utilized to explain the current $3.7\sigma$ derivation~\cite{Cherchiglia:2016eui,Endo:2017zrj,Cox:2018qyi,Kotlarski:2019muo}. The same contributions also arise for muon-electron scattering and therefore the MSSM is an excellent theory for us to demonstrate the argument presented in the previous section applied to a full model. For this purpose we use the framework of the previous section and compute the corrections from the muon vertex form factors~\eqref{eq:Vertex} and the photon propagator~\eqref{eq:SelfEnergy} arising from smuons and charginos. We neglect any corrections to the electron-photon vertex, since they are constrained by the electron $g-2$ and generally suppressed due to the small mass of the electron. 
We follow the notation set up in \cite{Martin1997,Lindner2018} with the chargino mass matrix defined as follows
\begin{subequations}
\begin{equation}
M_{\chi^0} = \begin{pmatrix}
\begin{matrix} M_1 & 0 \\ 0 &  M_2 \end{matrix} & M_{OD} \\ 
M_{OD}^T & \begin{matrix} 0& -\mu \\ -\mu & 0 \end{matrix} \end{pmatrix},
\label{neutralinomassmatrix} 
\end{equation}
\begin{equation}
M_{OD}=\begin{pmatrix}
 - c_b \, s_W\, M_Z &
s_b\, s_W \, M_Z \\
 c_b\, c_W\, M_Z & - s_b\, c_W \, M_Z \end{pmatrix},
\label{neutralinosubmatrix} 
\end{equation}
\begin{equation}
M_{\chi^\pm} = 
\begin{pmatrix}
 M_2 & \sqrt{2} s_b\, M_W\\
                              \sqrt{2} c_b\, M_W & \mu \\
                              \end{pmatrix},
\label{charginomassmatrix}
\end{equation}and
\begin{eqnarray}
M^2_{\tilde\mu}= \begin{pmatrix}
   m^2_{L,\mu} &
              - m_\mu \mu\tan\beta \\
 -m_\mu  \mu^*\tan\beta &
    m^2_{R,\mu}
\end{pmatrix},
\end{eqnarray}
\end{subequations}
 $m^2_{L,\mu}=m^2_{L} +(s_W^2 -\frac{1}{2})m_Z^2\cos\,2\beta, \, m^2_{R,\mu}= m^2_R -s_W^2 \, m_Z^2\cos\,  2\beta $ and the abbreviations $s_b=\sin \beta,\, c_b=\cos \beta, \, s_W=\sin \theta_W$ and $c_W=\cos \theta_W$. We neglect the soft SUSY breaking term $A_{\tilde{\mu}}$ in the smuon mass matrix, since it only amounts to small corrections in the smuon mass. The mass of the muon sneutrino is given through the left-handed smuon mass parameter
\begin{equation}
m_{\tilde{\nu}}^2=m_L^2+\frac{1}{2}M_Z^2\cos 2\beta \, .
\end{equation}  
The real and positive masses of the neutralinos, charginos and smuons can be found by diagonalizing the corresponding mass matrix
\begin{subequations}\label{Eq:neutralino}
\begin{eqnarray}
N^* M_{\chi^0} N^\dagger &=& {\rm diag}(
m_{\chi^0_1},
m_{\chi^0_2},
m_{\chi^0_3},
m_{\chi^0_4}),\\
U^* {M}_{\chi^\pm} V^\dagger &=& {\rm diag}(
m_{\chi^\pm_1},
m_{\chi^\pm_2}),\\
X M^2_{\tilde\mu}\, X^\dagger &=& 
{\rm diag}\, (m^2_{\tilde\mu_1}, m^2_{\tilde\mu_2}).
\end{eqnarray}
\end{subequations}
Under the assumption that the gaugino masses unify at some GUT scale
\begin{equation}
M_1=\frac{g_1^2 M_2}{g_2^2}\approx \frac{5}{3}  \tan^2 W \, M_2 \, ,
\end{equation}
we end up with five relevant MSSM parameters $M_2, \, \mu, \,\tan \beta, \, m_{L,\mu}$ and $m_{R,\mu}$.

The calculation of the relevant Feynman diagrams was performed in the following manner: the diagrams were generated with QGraf \cite{Nogueira:1991ex} and projected onto form factors defined in~\eqref{eq:SEFF} and \eqref{eq:VFF}. Scalar integrals were reduced to master integrals using integration-by-parts identities, generated by LiteRed \cite{Lee:2013mka}. The master integrals were then computed using QCDloop \cite{Carrazza2016}. Finally we renormalized our calculation in the on-shell scheme \cite{Hollik1999a}. 
\begin{table}
\begin{tabular}{|c|c|c|c|c|c|}
\hline
Set & $M_2$[GeV]& $\mu$[GeV] & $\tan \beta$ & $m_{L,\mu}$[GeV] & $m_{R,\mu}$[GeV] \\
\hline
\hline
$\text{L}$ & $200$ & $200$ & $4$ & $100$ & $100$  \\
\hline
$\text{H}\chi$ & $700$ & $700$ & $30$ & $300$ & $100$  \\
\hline
$\text{H}\tilde{\mu}$ & $200$ & $200$ & $30$ & $600$ & $600$  \\
\hline 
\end{tabular}
\caption{Definitions of parameter choices used for the MSSM calculation.}
\label{tab:MSSM}
\end{table}

\begin{figure}[t]
  \centering
  \includegraphics[scale=0.5]{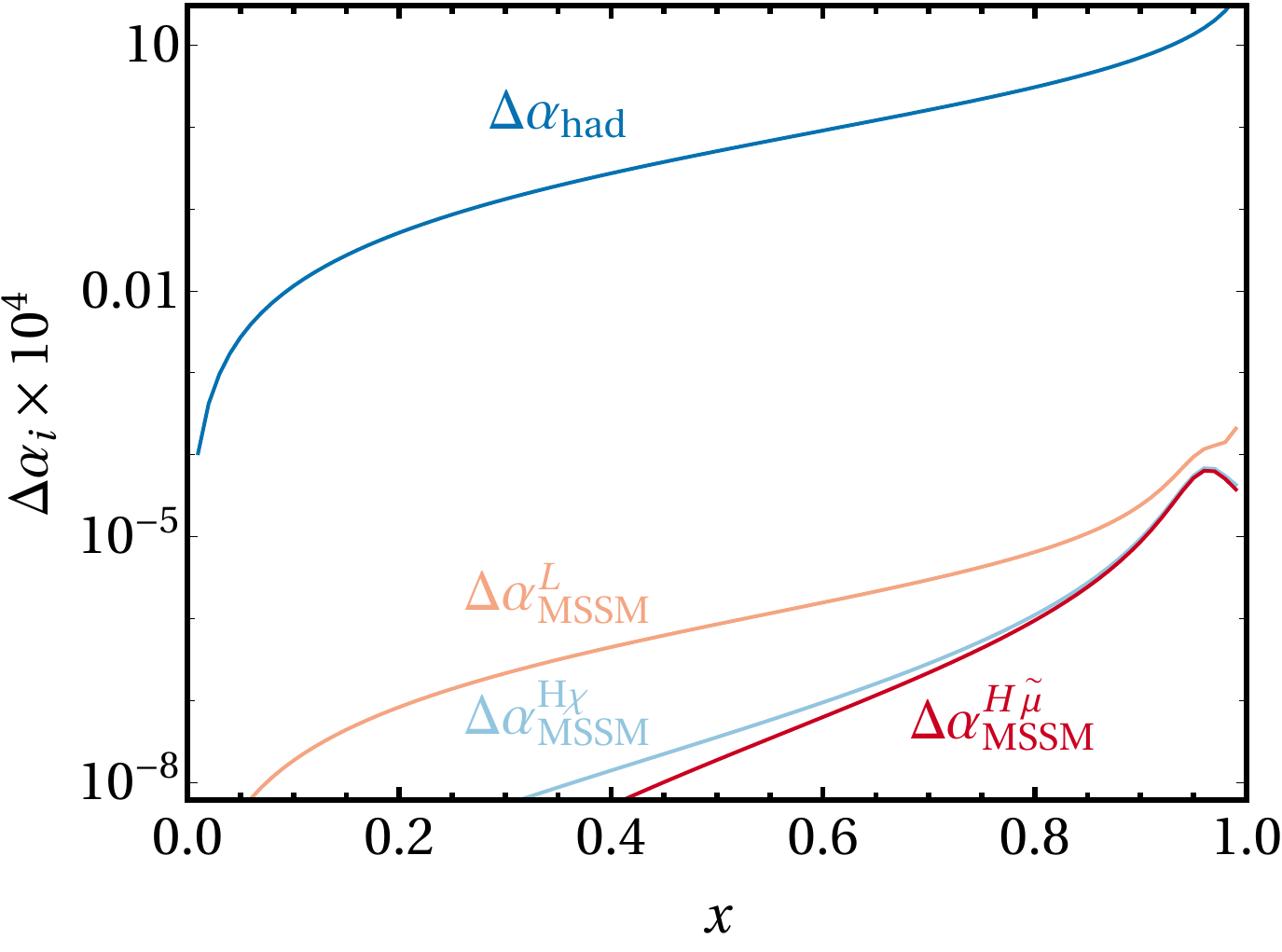}
  \caption{The plot shows the Leading Order hadronic contributions \ah~ in comparison with different parameter selections for the MSSM.}
  \label{fig:MSSMplot}
\end{figure}

Our results are shown in Fig.~\ref{fig:MSSMplot}, where we choose three sets of MSSM parameters as defined in Table~\ref{tab:MSSM}, which are compatible with the current $g-2$ discrepancy. The individual parameters are varied so as to preferentially select different loop diagrams. Regardless of the parameter choice we observe that the MSSM corrections are succifently small such as to effect $\delta a^{\rm{HLO}}_{\rm{BSM}}$ at the level of $\mathcal{O}(10^{-13})$ and therefore can be safely neglected. This validates with a specific model the more general statements made in the previous section regarding the overall impact of heavy loop induced new physics.   
 
\section{A $U(1)_{L_{\mu}-L_{\tau}}$ Model}

As a final example we consider a loop-induced BSM correction which corresponds to a dark photon gauged under the difference of muon and tau lepton numbers $L_{\mu}-L_{\tau}$~\cite{He:1990pn,He:1991qd}. Since there is no tree-level coupling to electrons this model is harder to constrain experimentally~\cite{Altmannshofer:2014pba} and thus still has an available window of parameter space compatible with $g-2$~\cite{Bauer2018}. 
For our discussion it represents an interesting test case, since the light mediating particle invalidates the argument given in the previous section regarding the smallness of self-energy style corrections (the kinematic suppression from the vertex diagrams is still present). These models therefore present a test of loop-induced BSM physics in a regime different from the MSSM example considered previously. In principle this model can be captured by an effective tree-level interaction by integrating out the loop, and thus could be constrained using the tree-level model (with a replacement $\epsilon^2\rightarrow \epsilon \epsilon'_{e}$). Here however we will use the full one-loop machinery to ensure a full comparison (also including vertex corrections) is made. 

\begin{figure}[t]
  \centering
  \includegraphics[scale=0.5]{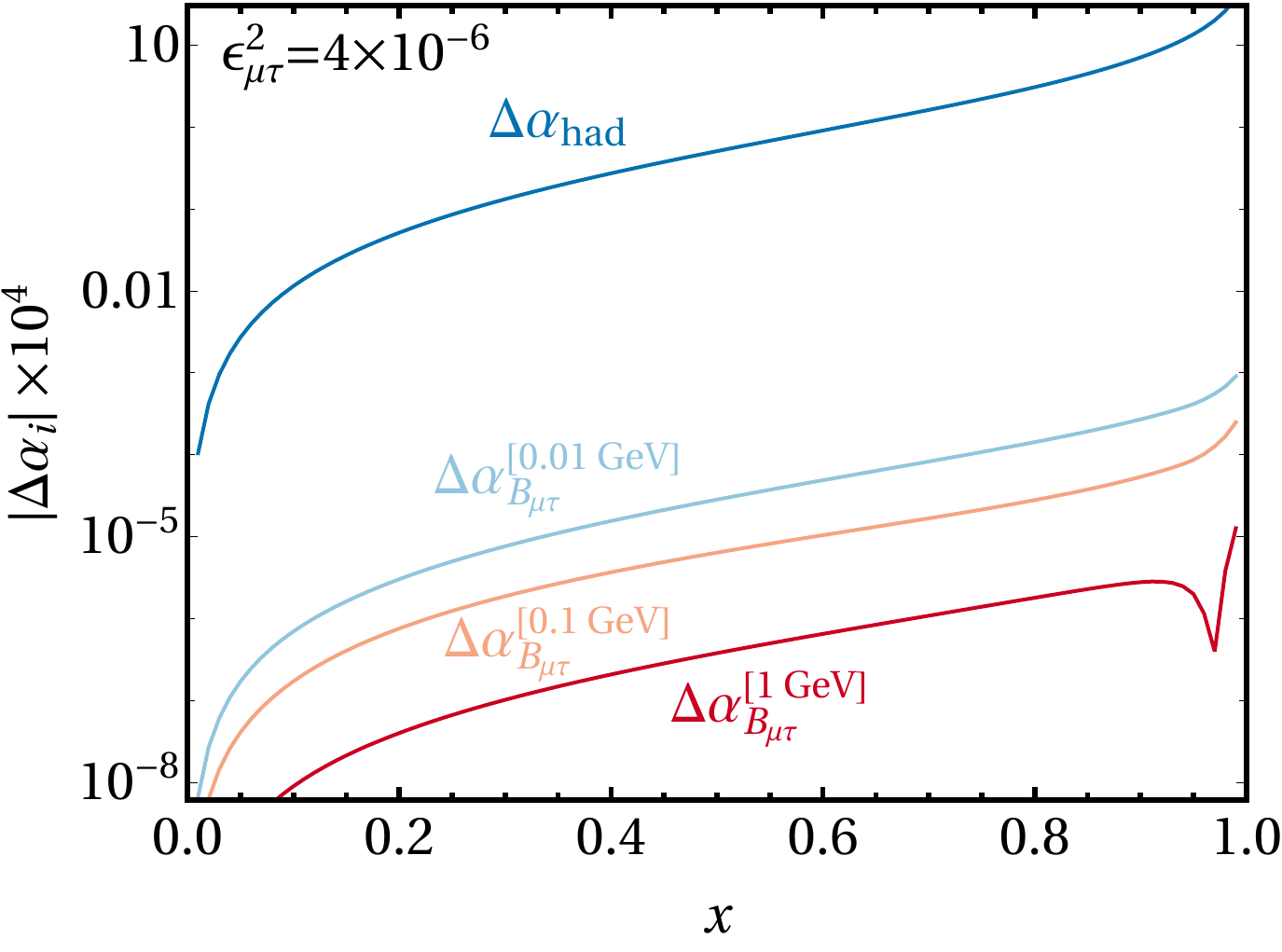}
  \caption{The plot shows the Leading Order hadronic contributions \ah~ in comparison with various $U(1)_{L_{\mu}-L_{\tau}}$ dark photon models $\Delta \alpha_{B_{\mu \tau}}^{[m_{A'}]}$.}
  \label{fig:Bmutauplot}
\end{figure}

Our results for this model are summarized in Fig.~\ref{fig:Bmutauplot}, where we plot $\Delta \alpha_{BSM}$ for three different choices of the dark photon mass (which would be broadly compatible with $g-2$). Although the contributions from the light new physics are considerably larger than those from the MSSM they are still suppressed to a percent level contribution to the error on \ah, and therefore can be neglected in the \muone~ analysis $(\delta a^{\rm{HLO}}_{\rm{BSM}} =\mathcal{O}(10^{-12}))$. From a theoretical viewpoint it is interesting to investigate the predictions in slightly more detail, as they have somewhat unique features compared to the other examples we have studied. For light mediators the dominant contribution comes from the electric form factor (e.g. the first term in ~\eqref{eq:VFF}), which is negative (for this reason we plot the absolute values in Fig.~\ref{fig:Bmutauplot}). On the other hand the contribution arising from the self-energy type corrections, has a positive sign. The two terms compete, and in particular at large $x$ the self-energy terms approach a constant value (for fixed coupling) regardless of the dark photon mass.  While the renormalized electric form factor acts like an effective coupling which decreases with increasing mass. 
As a result the shape of the vertex corrections as a function of $x$ is not sensitive to the mass of the dark photon. 
We note that magnetic form factor is subleading for all three choices such that total vertex correction is set by the electric form factor. 

For the two lighter cases studied ($m_{A'} = 0.01,0.1$ GeV) the electric form factor dominates over the entire $x$ range, and the prediction remains negative. 
For our heaviest case $m_{A}=1$ GeV the vertex suppression is sufficient that at larger values of $x$ the self-energy term dominates, as a result the 
prediction changes sign at large $x$.

\section{Conclusions}

A precision low-energy $\mu e$ scattering experiment offers the opportunity to perform an independent measurement of the LO hadronic running of $\alpha$  (\ah) with the possibility of producing a result with similar/improved uncertainties to existing calculations/extractions. Such a measurement would have an immediate application in the comparison between data and theory for the anomalous magnetic moment of the muon $g-2$. Currently there is significant tension (3.7 $\sigma$) between data and the predictions of the SM for this observable. Excitingly, this may be due to contributions, from as yet undiscovered BSM physics. New results for $g-2$ are expected from the Muon $g-2$ experiment at Fermilab this summer. 
Since $a^{\rm{HLO}}$~represents the second largest single contributor to the theoretical error budget, its extraction using an independent method, not plagued by low-energy hadron resonances (a problem for the current method relying on the optical theorem) is well motivated.  For this reason the \muone~ experiment has been proposed as a means of achieving this measurement through $t-$channel scattering of electrons and muons. 

The desired accuracy on the \muone~ experiment (in units of the anomalous magnetic moment of the muon) is about $2 \times 10^{-10}$, we studied the feasibility of obtaining this goal given the presence of unknown higher order corrections in the SM. While it is clear that a precision of 0.3\% on the fit is realistic given the proposed experimental methodology, there is an underlying dependence on missing terms in the theory which may alter the mathematical definition of the fitted function at a level much greater than this accuracy (i.e. one is not determining purely \ah~ but instead the combination of ~\ah+higher order perturbative terms). We conclude that the target of 10 ppm for the theoretical uncertainty is insufficient to obtain the desired accuracy on $a^{\rm{HLO}}$. We demonstrated this using an estimate of the order of magnitude of higher order corrections constructed from the leptonic running raised to the appropriate power, and by investigating powers of $\log(-t/m_e^2)$ which may enter into the perturbative expansion. Both analyses suggest that 
1 ppm is a more realistic target to achieve the $2\times 10^{-10}$ theoretical uncertainty. 

A putative BSM contribution to $g-2$ may be as large $\sim25 \times 10^{-10}$. One therefore should ask, if a BSM explanation is employed to address the $g-2$ discrepancy, what would its impact be on a similar scattering experiment? A natural worry is that BSM physics could be accidentally fitted into an extraction of \ah~ at the \muone~ experiment, and lead to a misinterpretation of $g-2$ data (in the worst case scenario, forcing artificial agreement with the ``SM''). It is, in our opinion, therefore crucial to understand  how different types of BSM signals would manifest themselves in the \muone~ experimental setup. Providing such a study was one of the principal aims of this paper. 
 
 Generic BSM physics capable of altering $\mu e$ scattering will  enter first at either tree- or loop-level and the potential impact in the two scenarios is rather different. For tree-level BSM physics the landscape is rather strongly constrained by previous collider experiments. This leaves two potential windows in the generic BSM parameter space, firstly the mediating BSM particle could be sufficiently heavy to have avoided direct production at LEP/LHC etc., or secondly the coupling of the mediating particle could be small enough that it is sufficiently weakly produced at colliders to have been observed. Heavy BSM  physics essentially replaces the tree-level diagram with a four-fermion vertex, and existing constraints are sufficient to render this of no concern to the \muone~ operational procedure. Lighter weakly coupled BSM physics is much more interesting from the \muone~ perspective. In particular dark photon models with an extended BSM matter sector might be able to make sizable contributions, while evading exclusion limits set by dark photon searches. Given the timescale of the experiment, and the current experimental interest in dark photons the available parameter space will be more tightly constrained by the time the \muone~experiment takes data. However, if a dark photon model is used to explain $g-2$, its impact on the \muone~ result should be computed to avoid double counting. 
  
Physics which enters first at loop-level is more subtle, we showed that in general there is a significant suppression from the kinematics and that generic models will contribute at a much smaller level than the anticipated error on \ah~. We demonstrated this with an explicit calculations in the MSSM or a dark photon arising from a gauged $U(1)_{L_{\mu}-L_{\tau}}$. 
 
 Given the importance of interpreting the $g-2$ difference as the breakdown of the SM we suggest that, if \ah~ information is used from $\mu e$ scattering, the BSM contribution to \ah~(as extracted from the data) is calculated for the model to ensure that the contribution is not large in both. Following the steps presented in this paper can provide an estimation of the size of the BSM physics, although a more rigorous analysis following the steps of the experiment should be conducted if warranted.

\section*{Acknowledgements} 
CW is supported by a National Science Foundation CAREER award number PHY-1652066.
US is supported by the National Science Foundation awards PHY-1719690 and PHY-1652066.
Support provided by the Center for Computational Research at the University at Buffalo.\\

\section*{Appendix A : Form of the Leptonic Running}
In this appendix we give the analytic expression for the leptonic running of $\alpha$ that was used to approximate the higher order corrections
\begin{widetext}
\begin{align}
\Delta \alpha_{\rm{lep}}&= \sum_{l=e,\mu\,\tau}\frac{\alpha}{\pi} \left(-\frac{5}{9}-\frac{4m_l^2}{3t} +\sqrt{1-\frac{4m_l^2}{t}} \, \frac{2m_l^2+t}{3t} \log\left( -\frac{1+\sqrt{1-\frac{4m_l^2}{t}}}{1-\sqrt{1-\frac{4m_l^2}{t}}} \right) \right) \, .
\end{align}
\end{widetext}

\bibliography{muone} 

\end{document}